\documentclass{emulateapj}
\usepackage{apjfonts}
\newcommand{\degs}{\mbox{$^{\circ}$}}

\shorttitle{Parallax and Proper Motion of PSR J0030+0451}
\shortauthors{Lommen et al.}

\begin{document}

\title{The Parallax and Proper Motion of PSR J0030+0451}

\author{Andrea N. Lommen and Richard A. Kipphorn}
\affil{Department of Physics and Astronomy, Franklin and Marshall College\\
501 Harrisburg Pike, Lancaster, Pennsylvania, 17603}

\author{David J. Nice}
\affil{Physics Department, Bryn Mawr College \\
Bryn Mawr, PA  19010}

\author{Eric M. Splaver}
\affil{Physics Department, Princeton University \\
Princeton, NJ  08544}

\author{Ingrid H. Stairs}
\affil{Department of Physics and Astronomy, University of British Columbia\\
6224 Agricultural Road, Vancouver, BC V6T 1Z1, Canada}

\and

\author{Donald C. Backer}
\affil{Department of Astronomy and Radio Astronomy Laboratory\\
University of California, Berkeley, CA 94720}

\submitted{Submitted to the Astrophysical Journal, 27 Sept 2005; revised 23 Dec 2005; accepted 6 Jan 2006}

\medskip

\begin{abstract}

We report the parallax and proper motion of millisecond pulsar J0030+0451,
one of thirteen known isolated millisecond pulsars in the disk of the Galaxy.
We obtained more than 6 years of monthly data 
from the 305 m Arecibo telescope at 430 MHz and 1410 MHz.
We measure the parallax of PSR J0030+0451 to be $3.3\pm0.9$\,mas, corresponding to a distance of
$300\pm90$\,pc.  
The Cordes and Lazio (2002) model of galactic electron
distribution yields a dispersion measure derived distance of 317 pc which
agrees with our measurement.
We place the pulsar's transverse space velocity in the range of 8 to 
17 km~s$^{-1}$, making this pulsar one of the slowest known.
We perform a brief census of velocities of isolated versus binary millisecond
pulsars.  We find the velocities of the two populations are indistinguishable.
However, the scale height of the binary population is twice that of the isolated population
and the luminosity functions of the two populations are different.  We suggest that the scale height difference
may be an artifact of the luminosity difference.

\end{abstract}


\section{Introduction}

A pulsar parallax can be combined with
a measurement
of the pulsar's dispersive delay (the Dispersion Measure or DM) to
provide an accurate measure of
the free electron density along the line of sight (LOS).  PSR J0030+0451 is one of fewer than a dozen
pulsars to have its parallax measured via timing
\citep{Kaspi94, Camilo94_1713, Sandhu97, Toscano99,
Wolszczan00, Jacoby03, Hotan04, Loehmer04, Splaver05}.  Another dozen have been measured
via VLBI (e.g. see \nocite{Brisken02, Chatterjee04} Brisken et al. 2002, Chatterjee et al. 2004, and also
this URL\footnote{
http://www.astro.cornell.edu/$\sim$shami/psrvlb/parallax.html}).
These measurements are important because
they give us most of our knowledge about
the galactic thermal electron distribution
\citep{Cordes02, Toscano99, Taylor93}.    

In addition, PSR J0030+0451 presents a rare evolutionary case as an isolated 
millisecond pulsar (MSP).
In the most popular MSP evolutionary model, MSPs
are formed via accretion
of matter from a companion star.  The incoming matter adds to the pulsar's angular
momentum, i.e., the pulsar is ``spun up.''  However, isolated MSPs present a conundrum:  they were presumably
spun up, 
yet they are without a companion which would have done so.  
One possible scenario is that the pulsar has ablated its companion \citep{Ruderman89}.

We expect MSPs to have lower velocities than the regular population, because the kick from
the supernova progenitor had to be small enough to leave the binary intact.  To produce
an isolated MSP, the binary must
remain intact during and after the supernova, but then after the spin-up phase the companion
must leave the system or be evaporated.
Several authors have debated whether isolated MSP velocities are lower, higher,
or indistinguishable from those of the general population of MSPs.  
\cite{McLaughlin04b} suggest we might expect isolated MSPs to have
higher velocities.  
They argue that if isolated MSPs are formed by ablation, we would expect them
to form from the tighter binaries which are more susceptible to ablation.
The correlation between tight binaries and higher velocities is suggested by \citet{Tauris96}.
\citet{McLaughlin04b} present the argument for faster velocities for
isolated MSPs as a counterpoint to 
their timing proper motion and scintillation measurements
which suggest the opposite, as do the measurements of 
\citet{Johnston98} and \cite{Toscano99}.  
\cite{Hobbs05}, however,
find the velocities of the populations to be indistinguishable.
We present a measurement of
the transverse velocity of PSR J0030+0451 which is unusually small, even compared to the
isolated MSP population.   We reconsider the question of the velocity
of isolated MSPs as compared to the binary MSP population.

Owing
to its small
timing residual, $\sim$1~$\mu$s, 
PSR J0030+0451 is a good candidate for membership in the Pulsar Timing Array (PTA), 
which is a collection
of pulsars that will be used for detecting gravitational radiation \citep{Jaffe03, Jenet04}.
For this reason, continued refinement of the timing model of
PSR J0030+0451 is important.
In fact, the PTA, as it is conceived, is an interferometer, so
a variety of baselines will be important to its operation.
PSR J0030+0451 may
be particularly useful in this regard because
it has large angular separation from PSRs B1855+09, J1713+07 and
J0437-4715, which 
are among the most stable and precise pulsars
\citep{Kaspi94, vanStraten01, Lommenthesis}.

In \S2 we present a significant refinement to the timing model previously published
\citep{Lommen00}.
We discuss our measurements of parallax and proper motion in \S\ref{sec:pm_px}. 
We include the effects of the solar 
wind in our analysis, which we discuss in \S4.
In \S5,\S6, and \S7 we discuss the space velocity of J0030+0451, corrections to its measured period
derivative, and the implications of its measured distance for the local interstellar medium (LISM).
In \S8 we summarize our conclusions.

\section{Arecibo Observation and Data Reduction}

We have conducted timing observations
over a 6.5 year period, from 1997 December to 2004 July, at the Arecibo Observatory, using the Arecibo-Berkeley 
Pulsar Processor (ABPP) and the Princeton Mark~IV system.  
Details of ABPP signal processing can be found in 
\citet{Lommen00}, where
the first three years of these data are presented.  Details of Mark~IV signal processing can be found
in \citet{Stairs00}.  

Over the course of each observation, incoming signals from two
orthogonal polarizations were coherently dedispersed, squared to
obtain power measurements, and folded modulo the pulse period for
intervals of three minutes.  Opposite polarizations were summed, using
absolute flux calibrations of the receivers whenever possible or, when
absolute flux calibration was not available, using the system
temperature of the pulsed noise source as published by telescope
operations.

Pulse times of arrival (TOAs) were derived from the calibrated three
minute integrations using conventional algorithms.  The TOAs from a
given frequency and instrument on a given day were averaged into a
single effective TOA for that day, the uncertainty of which was
estimated from the spread of the individual three-minute TOAs.

\begin{figure}
\plotone{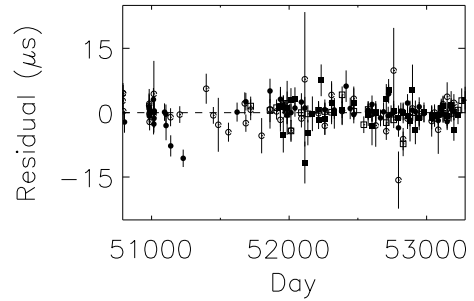}
\caption{
Residual vs day for the model presented in Table \ref{tab:0030}.  Circles are ABPP data.  Squares
are Mark IV data.  Filled points are 1410 MHz.  Open points are 430 MHz data.
\label{fig:resid}
}   
\end{figure}

We performed a weighted fit to the averaged TOAs using 
{\sc tempo}\footnote{See http://pulsar.princeton.edu/tempo}. 
We used ecliptic coordinates to minimize covariance between
the two components of the position and proper motion measurements.  This
is usually necessary when
the ecliptic latitude is low (1.44\degs\,in our case).
Table 1 
shows updated spin, astrometric, and other parameters for PSR J0030+0451.  The 
root mean squared (rms) of the timing residuals quoted in 
Table 1, is calculated using TOAs
from profiles averaged over 30 
minutes.  We estimated uncertainties in 
all the fitted parameters by doubling errors given by 
{\sc tempo}, an ad hoc procedure that attempts to account for timing noise and other possible systematic errors.
The quoted uncertainty in parallax also includes a term added in quadrature (0.3 mas) due to the solar
wind (see \S \ref{sec:electron}).
Residual arrival times after subtracting the best fit are shown in Figure \ref{fig:resid}.  

\begin{deluxetable*}{ll}
\tablecolumns{2}
\tablewidth{0pc}
\tablecaption{Parameters for pulsar J0030+0451
\label{tab:0030}
}
\tablehead{
\colhead{Parameter } & \colhead{Value\tablenotemark{a}}
}
\startdata
Ecliptic longitude (deg)      \dotfill               & 8.91036695(6) \\ 
Ecliptic latitude (deg)   \dotfill                   & 1.445692(3) \\
Period (s) \dotfill                                  & 0.00486545320829334(3) \\ 
Period derivative (s\,s$^{-1}$) \dotfill              & 1.0162(1)$\times10^{-20}$ \\ 
Proper motion in ecliptic longitude (mas yr$^{-1}$)  \dotfill                     & $-$5.74(9) \\ 
Absolute value of proper motion in ecliptic latitude (mas yr$^{-1}$)   \dots                      & $< 10$ \\ 
Dispersion measure (pc cm$^{-3}$) \dotfill           & 4.3328(2) \\ 
Epoch (MJD)  \dotfill                                & 52035 \\ 
Solar {\em n}$_{o}$ (e$^{-}$ cm$^{-3}$)    \dotfill        & 6.9(2.1) \\
Solar {\em $\dot{n}$}$_{o}$ (e$^{-}$cm$^{-3}$yr$^{-1}$) \dotfill& 0.0(3) \\
Number of epochs of data \tablenotemark{b} \dotfill  & 56 \\ 
Timing data span (MJD)     \dotfill                  & 50,790~$-$~53,280 \\
Right ascension(J2000)\dotfill                       & 00 30 27.4308(6) 	\\ 
Declination(J2000)            \dotfill               & +04 51 39.72(1) \\ 
Galactic longitude (deg)              \dotfill       & 113.141135(1) \\ 
Galactic latitude (deg) \dotfill                     & $-$57.611237(3) \\ 
DM derived distance (pc)\tablenotemark{c} \dotfill   & 317(25) \\
Characteristic age (yr)  \dotfill                    & $7.8\times10^{9}$ \\ 
Magnetic field strength (G)\tablenotemark{d} \dotfill& $2.7\times10^8$ \\ 
Column electron density along LOS (e$^{-}$ cm$^{-3}$)\dotfill         & 0.014(2) \\ 
Parallax (mas)   \dotfill                            & 3.3(9) \\ 
Parallax derived distance (pc)	\dotfill   & 300(90) \\ 
Magnitude of transverse velocity (km~s$^{-1}$) 	\dotfill	  & 8$-$17 \\ 
RMS residual at 430 MHz ($\mu$s, ABPP/MarkIV) \dotfill & 2.3/1.0 \\
RMS residual at 1410 MHz ($\mu$s, ABPP/MarkIV) \dotfill & 2.2/2.2 \\
\enddata
\tablenotetext{a} {Uncertainties in parentheses refer to the last digit quoted.
Note that a $n_0$ of 6.9 cm$^{-3}$ was assumed in all fits.}
\tablenotetext{b} {We define ``epoch" as data separated by 2 weeks or more.}
\tablenotetext{c} {Model from NE2001.}
\tablenotetext{d}  {B$_{0}$ = 3.2 x 10$^{19}$ G[P(s)P$_0$]$^{1/2}$.}
\end{deluxetable*}

We examined timing solutions in which DM was allowed to vary over time, and we found that they did not significantly improve the fit quality or change the timing model parameters.
However, it is interesting to compare our upper limit
on $\dot{\rm DM}$, the time derivative of DM, to  
the correlation between DM and $\dot{\rm DM}$ measured
by \citet{Backer93}.  
When $\dot{\rm DM}$ is included in the timing model, its best-fit value is
$(1.9 \pm 1.8) \times 10^{-5}$ cm$^{-3}$pc~yr$^{-1}$, which implies an upper limit (with 95\% confidence) of
$5.5 \times 10^{-5}$ 
cm$^{-3}$pc~yr$^{-1}$. This value is smaller
than expected according to \citet{Backer93} by about a factor of 5.
If more small-DM small-$\dot{\rm DM}$ pulsars are found it might suggest
that $\dot{\rm DM}$ is a stronger function of DM than we would expect
for wedgelike thermal plasma perturbations distributed randomly along
the LOS.

\section{Proper Motion and Parallax}
\label{sec:pm_px}

\begin{figure*}
\plottwo{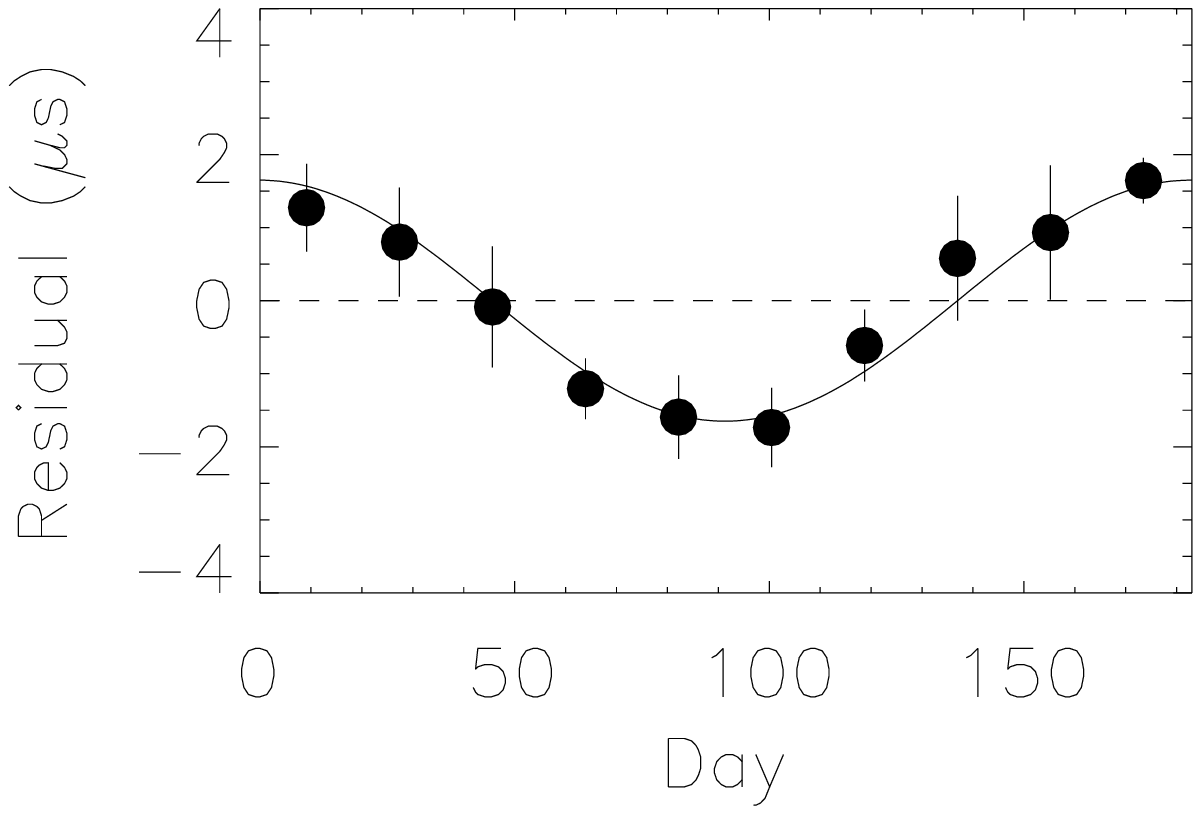}{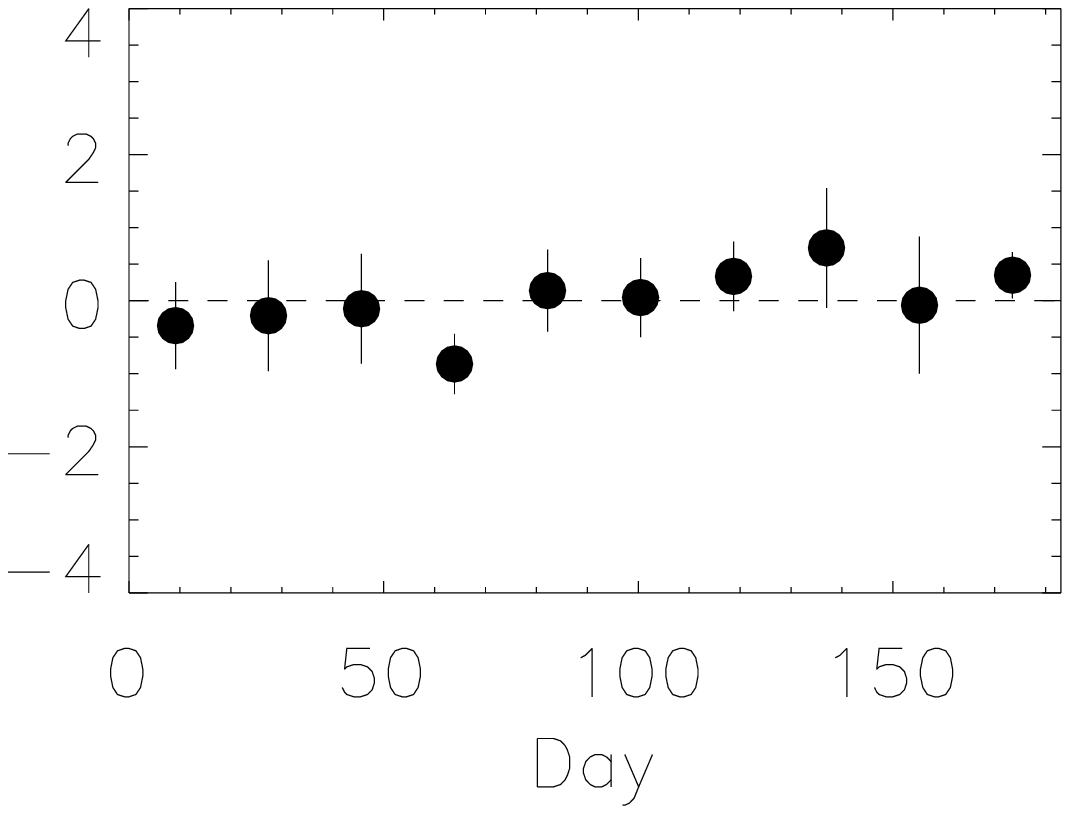}
\caption[FIG. 1.-Timing residuals for PSR J0030+0451]{
Timing residuals for J0030+0451 folded over a period of a half year and
binned every 18 days. Left: Residuals with no parallax
fit.  Right:  Residuals after removal of parallax of 3.3 mas from
the data in the left hand panel.
The curve that represents the 3.3 mas parallax is
shown in the left hand panel.  Day 0 is when the Earth-Sun-pulsar angle is 
90\degs.
\label{fig:parallax}
}   
\end{figure*}

Evidence for a significant parallax measurement is shown in
Figure \ref{fig:parallax}, which compares residuals of the pulse arrival
times with and without parallax incorporated into the
timing model.
The figure
shows averaged timing residuals versus day number binned in increments of 18 days  
and folded over a half-year period to provide the best visibility for the parallax signature.
The data have been fit for 
proper motion but not for parallax in the left part of Figure \ref{fig:parallax}.  Data on the right
have been fit for both proper motion and for parallax.  
We have superimposed the
best-fit parallax curve onto the pre-fit data.  
On close examination of Figure \ref{fig:parallax}, there is a subtle difference in
the positions of the data points relative to the fit curve (left plot) and
the horizontal axis (right plot).
This is not an unexpected result of (a) a global fit to
all the parameters with and without parallax and (b) the post-fit binning.

Our best-fit value for parallax, $\pi=3.3\pm 0.9$\,mas, 
includes both measurement uncertainty of 0.88 mas and systematic
uncertainty
due to the solar wind model, 0.3\,mas, which will be discussed in the following section. 

We measure a proper motion of  $-5.74\pm 0.09$\,mas yr$^{-1}$ in the plane of the
ecliptic.
Proper motion out of the ecliptic plane 
is naturally difficult to measure given the pulsar's position, a problem which was
compounded by its high covariance with variations in dispersion measure. 
We
quote a conservative upper limit on the magnitude of the out-of-plane proper motion
of 10 mas~yr$^{-1}$.

We can combine
our distance measurement with our proper motion measurements
to yield the pulsar's velocity.
We obtain a velocity of $-8.3\pm 0.1$~km~s$^{-1}$ in the plane of the
ecliptic and an upper limit of $14.4$~km~s$^{-1}$ out of the plane
of the ecliptic.
Thus, the pulsar's transverse velocity lies between 8 and 17 km~s$^{-1}$.
This confirms the value
presented by \citet{Nicastro01}, 
$9\pm6$\,km~s$^{-1}$, 
which they found using scintillation measurements\nocite{Nicastro01}.  

The measured proper motion of PSR J0030+0451 results not just from its
motion relative to its local standard of rest (LSR), but also from the
difference between its LSR and the LSR of the sun and from the solar
motion.  After considering these effects (see \S 5 for details), we
calculate its transverse velocity relative to its LSR to be between 4
and 20 km\,s$^{-1}$.

This is one of the slowest transverse velocities measured for any
pulsar.  Young pulsars (those which have not been spun up) have a mean
velocity of 265 km~s$^{-1}$ \citep{Hobbs05}, a factor of three
times faster than MSPs (\S 5;  see also Hobbs et al. 2005; Cordes \&
Chernoff 1997; Nice \& Taylor 1995).  PSR J0030+0451 is, in fact, only
one tenth as fast as the average MSP (\S 5).
\nocite{Hobbs05, Cordes97, NiceTaylor95}

\section{The Solar System Electron Density}
\label{sec:electron}

PSR J0030+0451 is roughly in the ecliptic plane (ecliptic latitude is 1.44\degs) 
where the solar system provides additional dispersion caused by the charged particles
of the solar wind.  \cite{Issautier01} show that the solar electron density,
{\em n}$_{e}$, can be modeled as $n_0/r^{2}$, where $r$ is the 
radial distance to the sun in astronomical units, 
and $n_0$ is roughly 10 cm$^{-3}$  
\citep{Splaver05, Issautier01}.
This simple model is the default in {\sc tempo}.  Maintaining the $1/r^{2}$ dependence, we searched
$\chi^2$ space for the best value of $n_0$, which we found to be $6.9\pm2.1$\,cm$^{-3}$.

Reducing $n_0$ from 9.0 to 6.9 cm$^{-3}$ reduces the fitted parallax by about 0.3 mas which 
increases the distance by about 30 pc.  These two parameters are covariant 
because the model of the solar wind
yields an annual pattern of pulse delays with a strong cusp when the
pulsar is behind the sun with respect to the earth.  A Fourier
decomposition of this delay pattern yields significant terms with
annual periodicity (covariant with pulsar position), semi-annual
periodicity (covariant with parallax), as well as higher order terms.
The additional 0.3 mas uncertainty in parallax due to the solar wind was added in 
quadrature to the doubled {\sc tempo} error. 

Note that in the case of PSR J1713+0747, another ecliptic plane MSP, \cite{Splaver05}
found it necessary to eliminate data within 30\degs\, of the sun.  In our case we did not find that
this significantly improved the fit, nor did it substantially change any fitted parameters.

We wondered if it would be possible to measure a change in the solar wind on its 11-year cycle. 
Using {\sc tempo} we mapped $\chi^2$ space in the range of 
$-$10 cm$^{-3}$~yr$^{-1} < \dot{n}_o < 10$ cm$^{-3}$~yr$^{-1}$ 
but found that the minimum in
$\chi^2$ occurred at an essentially null value of $0.0\pm0.3$\,cm$^{-3}$~yr$^{-1}$,
indicating that the change is currently beyond our measurement capability.

\section{Velocities of MSPs:  isolated vs binary}

\cite{Hobbs05} presents an extensive study of the velocities of various
sub-groups of pulsars, including isolated and binary recycled pulsars.
They find that isolated recycled pulsars are not significantly slower 
($77 \pm 16$\,km~s$^{-1}$)
than binary recycled pulsars ($89 \pm 15$\,km~s$^{-1}$).
When the velocity of PSR J0030+0451 (using 9 km~s$^-1$)
is added to the sample the average
velocity of isolated recycled pulsars becomes $68 \pm 16$\,km~s$^{-1}$,
which is still not significantly different from the average of the
binary recycled pulsars.
In contrast, both \cite{Johnston98} and \cite{Toscano99} previously claimed to see evidence that
isolated MSPs are slower than binary MSPs. 
However, when one uses a more recent Galactic electron density model
to estimate the distances to the pulsars \citep{Cordes02}, the
discrepancy disappears.
This is a potent reminder that a distance
model has a significant effect on conclusions drawn from it.
\cite{McLaughlin04b} also find the isolated MSP population to have
a slightly lower average velocity (70 km~s$^{-1}$) as compared to the
binary population. They caution against drawing conclusions
from velocity data which are based on imprecise distances.

We performed an analysis of the velocity data that is
slightly different from
the analysis done by \citet{Hobbs05} with
similar results.  
We defined an MSP to be any pulsar with period $P< 0.01$\,s, which
provides a narrower sample than that used by
\cite{Hobbs05}
who defined MSPs to be any pulsar with period $P<0.1$\,s and period
derivative  $\dot{P}<10^{-17}$. 
Our criterion means that 
the binary pulsars in our sample nearly all have helium white dwarf
companions, whereas Hobbs et al. used binaries with
a mix of companion types.

The 29 MSPs in the Galactic Disk with measured
proper motions are shown in
Table \ref{tab:velocities}.
PSR J1730-2304, which has no measured declination proper motion, has
been included in the table for completeness but has not been included in
any of the following calculations.  We corrected each pulsar's velocity to its
LSR as follows.  
We used the measured proper motion and distance to calculate a three
dimensional vector representing the (two dimensional) transverse
motion of the pulsar in the reference frame of the Sun.  We then
removed the solar motion and rotated the resulting vector from the LSR
of the sun to the LSR of the pulsar.  Finally, we recovered those two
components of the vector which are perpendicular to the line of sight.
This computation required selecting a value for the unknown
LOS velocity of the pulsar;  we chose a value appropriate
for a star at rest in the pulsar's LSR.

\begin{deluxetable*}{r@{}lr@{}llr@{}llrrl}
\tabletypesize{\scriptsize}
\tablecolumns{7}
\tablewidth{0pc}
\tablecaption{
Velocities of Millisecond Pulsars in the Galactic Disk
\label{tab:velocities}
}
\tablehead{
\colhead{ } & \colhead{Pulsar} & \multicolumn{3}{c}{$\mu_\alpha$}
& \multicolumn{3}{c}{$\mu_\delta$}
& \colhead{Distance} & \colhead{$v_t$}
& \colhead{Reference}
\\
\colhead{} & \colhead{ } & \multicolumn{3}{c}{(mas~yr$^{-1}$)}
& \multicolumn{3}{c}{(mas~yr$^{-1}$)}
& \colhead{(pc)} & \colhead{(km~s$^{-1}$)}
& 
}
\startdata   
\cutinhead{Isolated MSPs}
J & 0030+0451 & $\mu_\lambda = -5$ &.84 &$\pm$\phn0.09 & $|\mu_\beta| < 10$&& &  310\tablenotemark{p} &   $<$20 & This work \\
J & 0711$-$6830 &    $-$15&.7 &$\pm$\phn0.5 &     15&.3 &$\pm$\phn0.6  &  860\tablenotemark{n} &  113 & \cite{Toscano99a}\\
J & 1024$-$0719 &      $-$41&&$\pm$\phn2 &     $-$70& &$\pm$\phn3  &  200\tablenotemark{o} & 70 & \cite{Toscano99a}\\
J & 1730$-$2304 &     20&.5 &$\pm$\phn0.4 &   \multicolumn{2}{c}{\nodata}&&  510\tablenotemark{n} &  $>$50 & \cite{Toscano99a}\\
J & 1744$-$1134 &    18&.64 &$\pm$\phn0.08 &    $-$10&.3 &$\pm$\phn0.5  &  360\tablenotemark{p} &  31 & \cite{Toscano99a}\\
  B & 1937+21 &    $-$0&.130 &$\pm$\phn0.008 &    $-$0&.469 &$\pm$\phn0.009 & 3600\tablenotemark{n} &  87 & \cite{Kaspi94}\\
J & 1944+0907 &     12&.0 &$\pm$\phn0.7 &      $-$18& &$\pm$\phn3  & 1800\tablenotemark{n} & 173 & \cite{Champion05}\\
J & 2124$-$3358 &      $-$14&&$\pm$\phn1 &      $-$47& &$\pm$\phn1  &  270\tablenotemark{n} &  48 & \cite{Toscano99a}\\
J & 2322+2057 &      $-$17&&$\pm$\phn2 &      $-$18& &$\pm$\phn3  &  790\tablenotemark{n} &  79 & \cite{NiceTaylor95}\\
\cutinhead{Binary MSPs}
J & 0437$-$4715 &  121&.438 &$\pm$\phn0.006 &      $-$71&.438 &$\pm$\phn0.007  &  140\tablenotemark{p} &  84 & \cite{vanStraten01}\\
J & 0613$-$0200 &      2&.0 &$\pm$\phn0.4 &       $-$7& &$\pm$\phn1  & 1700\tablenotemark{n} &  60 & \cite{Toscano99a}\\
J & 0751+1807	&$\mu_\lambda = 0$&.35&$\pm$\phn0.03& $\mu_{\beta}=-6$&&$\pm$\phn 2 	& 1150\tablenotemark{n}	&  22	& \cite{Nice05}\\
J & 1012+5307 &      2&.4 &$\pm$\phn0.2 &    $-$25&.2 &$\pm$\phn0.2  &  840\tablenotemark{o} &  107 & \cite{Lange01}\\
J & 1045$-$4509 &       $-$5&&$\pm$\phn2 &        6& &$\pm$\phn1  & 1940\tablenotemark{n} &  119 & \cite{Toscano99a}\\
J & 1455$-$3330 &        5&&$\pm$\phn6 &       24& &$\pm$12  &  530\tablenotemark{n} &  71 & \cite{Toscano99a}\\
J & 1640+2224 &     1&.66 &$\pm$\phn0.12 &    $-$11&.3 &$\pm$\phn0.2  & 1160\tablenotemark{n} &  67 & \cite{Loehmer05}\\
J & 1643$-$1224 &        3&&$\pm$\phn1 &       $-$8& &$\pm$\phn5  & 2320\tablenotemark{n} &  96 & \cite{Toscano99a}\\
J & 1709+2313 &     $-$3&.2 &$\pm$\phn0.7 &     $-$9&.7 &$\pm$\phn0.9  & 1390\tablenotemark{n} &  57 & \cite{Lewandowski04}\\
J & 1713+0747 &      4&.917 &$\pm$\phn0.004 &     $-$3&.933 &$\pm$\phn0.010 & 1100\tablenotemark{p} &  30 & \cite{Splaver05}\\
B & 1855+09 &    $-$2&.94 &$\pm$\phn0.04 &    $-$5&.41 &$\pm$\phn0.06 &  910\tablenotemark{p} &  17 & \cite{Kaspi94}\\
J & 1909$-$3744 &     $-$9&.6 &$\pm$\phn0.2  &    $-$35&.6 &$\pm$\phn0.7  &  820\tablenotemark{p} & 131 & \cite{Jacoby03}\\
J & 1911$-$1114 &       $-$6&&$\pm$\phn4 &      $-$23& &$\pm$13  & 1220\tablenotemark{n} & 128 & \cite{Toscano99a}\\
B & 1953+29 &     $-$1&.0 &$\pm$\phn0.3 &     $-$3&.7 &$\pm$\phn0.3  & 4610\tablenotemark{n} &  128 & \cite{Wolszczan00_4msp}\\
B & 1957+20 &    $-$16&.0 &$\pm$\phn0.5 &    $-$25&.8 &$\pm$\phn0.6  & 2490\tablenotemark{n} & 325 & \cite{Arzoumanian94}\\
J & 2019+2425 &    $-$9&.41 &$\pm$\phn0.12 &    $-$20&.60 &$\pm$\phn0.15  &  1490\tablenotemark{n} &  142 & \cite{Nice01}\\
J & 2051$-$0827 &       1& &$\pm$\phn2 &      $-$5& &$\pm$\phn3  & 1040\tablenotemark{n} &  42 & \cite{Stappers98}\\
J & 2129$-$5721 &        7&&$\pm$\phn2 &       $-$4& &$\pm$\phn3  & 1340\tablenotemark{n} &  48 & \cite{Toscano99a}\\
J & 2229+2643 &      1& &$\pm$\phn4 &      $-$17& &$\pm$\phn4  & 1440\tablenotemark{n} & 130 & \cite{Wolszczan00_4msp}\\
J & 2317+1439 &     $-$1&.7 &$\pm$\phn1.5 &      7&.4 &$\pm$\phn3.1  & 820\tablenotemark{n} &  20 & \cite{Camilo96}\\
\enddata
\tablenotetext{p} {Distance from parallax.}
\tablenotetext{n} {DM distance from NE2001.}
\tablenotetext{o} {Some other method used to acquire distance.  The text of the cited reference should
be consulted for details.}  
\end{deluxetable*}

This corrected
velocity is listed in the second to last column of Table 3.
The
average corrected velocity of the isolated MSPs is $86 \pm 19$ km~s$^{-1}$ whereas
the average corrected velocity of all the binary MSPs is $91 \pm 28$ km~s$^{-1}$.  
(The uncorrected averages are 79 and 90 km~s$^{-1}$ respectively).
If one
allows the sample to include only those proper motions which have been measured
to better than 2$\sigma$ the average corrected velocities are
$86 \pm 19$\,km~s$^{-1}$ and $99 \pm 33$\, km~s$^{-1}$ respectively.  (Uncorrected
averages are 79 and 99 km~s$^{-1}$.)  In each case
the isolated MSP population is indistinguishable from the binary MSP population.
The 2$\sigma$ cutoff in velocity introduces a selection of higher velocity
pulsars. Thus, the average velocities are higher in that case. 

\begin{figure}
\plotone{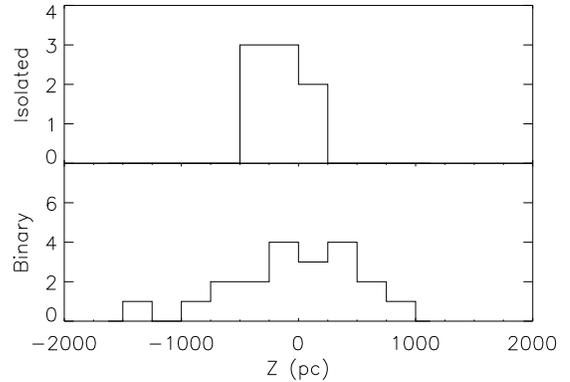}
\caption{
Histograms of height above the galactic plane for the isolated MSP 
population (upper)
and the binary MSP population (lower).
\label{fig:zheight}
}   
\end{figure}

An alternative statistic for evaluating the dynamics of pulsar populations
is the distribution of heights above
or below the galactic plane, $z$.  For the pulsars listed, one finds 
that the standard deviation from zero for the binary MSP population
is twice that of the isolated MSP population:  $570 \pm 90$\,pc vs 
$280 \pm 65$\,pc.
Figure \ref{fig:zheight} shows a histogram of $z$ for each population.  The
isolated MSP population is represented in the upper half of the figure,
the binary MSP population in the lower half. 

Figure \ref{fig:zheight} 
shows that the known isolated MSPs are closer to the Plane than
are the known binary MSPs.  This could be either a reflection of
differences in the intrinsic spatial distributions of the two types of
MSPs, or a selection effect.  A smaller intrinsic spread in scale
heights for isolated MSPs is only possible if that population also has a
smaller intrinsic velocity distribution, so that the objects do not
travel as far from the Plane as they oscillate in the Galaxy's
potential.  Our determination that the two velocity distributions are in
fact indistinguishable makes this scenario unlikely.  However, with
identical velocity distributions, a difference in intrinsic {\it
luminosity} distributions would cause the less-luminous population to be
detected only to smaller distances and hence only to smaller scale
heights.  In fact, \citet{Bailes97} find that luminosities of
isolated and binary MSPs are different at the 99.5\% confidence level,
with the isolated MSPs being intrinsically dimmer.  We have confirmed
their results with an updated catalog; also, a simple examination of the
median distance of the isolated population (510 pc) compared to
the median distance of the binary population (1155 pc) 
suggests that the
isolated MSPs must be less luminous.

\section{Corrections to $\dot{P}$}

Using our upper limit on the proper motion, we can calculate
the upper limit of the Shklovskii correction to the period derivative $\dot{P}$.  We find that
$4.4 \times 10^{-22}$
or not quite 5\% of
the measured $\dot{P}$ may be due to proper motion. 
The acceleration toward the disk of the Galaxy is about
the same size in the other direction,
$-5.0 \times 10^{-22}$.
The acceleration in the disk makes a much smaller contribution, $-2.2 \times 10^{-23}$.
Combining these corrections changes the measured $\dot{P}$ by less than 1\%.
 
\section{The Local Interstellar Medium}

The parallax measurement of $3.3 \pm 0.9 $\,mas yields a distance of $300 \pm 90$~pc.  This
falls in between the distance estimates made by the old (Taylor \& Cordes 1993, hereafter TC93) 
\nocite{Taylor93} and new (Cordes
\& Lazio 2002, hereafter NE2001)\nocite{Cordes02}
distance models:  230 and 317 respectively.  Both estimates are based on a 
DM of $4.3328\pm0.00020$\,cm$^{-3}$pc \citep{Lommen00}.
In the second galactic quadrant, between 
90$^{\circ}$ and 180$^{\circ}$ galactic longitude, PSR J0030+0451 is the first pulsar with a measured parallax,
so it provides an important check of the NE2001 model.  
In addition to a parameterized model of the LISM,
the model adds a number of clumps and voids to the previous TC93 model, but nothing
in the direction of PSR J0030+0451.   The agreement of our distance with theirs suggests
no significant clumps or voids exist along this LOS. 

Nearby pulsars with known parallaxes are very useful for studying the LISM.
Scintillation parameters measure density fluctuations and
have been used to map out the LISM, but
are unable to measure densities.  The density measurements must come
from parallax
\citep{Bhat98}.  These authors model the LISM explicitly as a low density   
bubble surrounded by a shell of much higher density fluctuations, but
more pulsars with known distances are required to confirm the model.
PSR J0030+0451 will therefore be a marvelous tool for studying the LISM.

\section{Conclusions}

We have measured the parallax of PSR J0030+0451
to be $3.3\pm0.5$\,mas.
We have measured its proper motion to be $-5.74\pm 0.09$\,~mas~yr$^{-1}$ in the plane of the
ecliptic and have established an upper limit on its motion out of the plane at 
10 mas~yr$^{-1}$.
The is one of the lowest velocities
measured for any pulsar and is noteworthy even within the relatively
low-velocity millisecond pulsar population.

Combining proper motion data from this pulsar with the collection of existing MSP proper
motion measurements, we find the statistical
properties of the transverse velocities of isolated and binary MSPs are indistinguishable from each other.
We do, however, find that the average $z$-height of the isolated MSPs is half that of the binary MSPs.
We suggest that a luminosity difference between the two classes of objects, such as
that suggested by \citet{Bailes97}, \citet{Kramer98}, and \citet{Hobbs04}, is the simplest
way to account for both the observed
difference in $z$-height and the similarity of the velocity distributions.

\acknowledgments
We are grateful to the Arecibo telescope operators.
We thank
Robert Ferdman, Paul Demorest, Paulo Freire, Duncan Lorimer, Ramachandran, and Kiriaki Xilouris,
for valuable discussions and for assisting with observations.  
We thank the referee, Simon Johnston, for substantial comments that significantly
improved the manuscript.
The Arecibo Observatory is a facility of the National
Astronomy and Ionosphere Center, operated by Cornell University under 
a cooperative agreement with the National Science
Foundation (NSF).   
ANL acknowledges a Research Corporation award in support of this research.  
DJN is supported by NSF grant AST-0206205. 
IHS holds an NSERC UFA and is supported by a Discovery Grant.
DCB acknowledges support from NSF AST-9987278 for ABPP instrumentation and
NSF AST-0206044 for the science program.

\end{document}